# Evaluating the performance of ionic liquid coatings for mitigation of spacecraft surface charges

**Mirco Wendt[1], Regina Lange[1], Franziska Dorn[1], Jens Berdermann[2], Ingo Barke[1] and Sylvia Speller[1]**

[1]University of Rostock, Institute for Physics
Albert-Einstein-Str. 23, 18106 Rostock, Germany

[2]German Aerospace Centre (DLR), Institute for Solar-Terrestrial Physics
Kalkhorstweg 53, 17235 Neustrelitz, Germany

**Abstract:** To reduce the impact of charging effects on satellites, cheap and lightweight conductive coatings are desirable. We mimic space-like charging environments in ultra-high vacuum (UHV) chambers during deposition of charges via the electron beam of a scanning electron microscope (SEM). We use the charge induced signatures in SEM images of a thin ionic liquid (IL) film on insulating surfaces such as glass, to assess the general performance of such coatings. In order to get a reference structure in SEM, the samples were structured by nanosphere lithography and coated with IL. The IL film (we choose BMP DCA, due to its beneficial physical properties) was applied *ex situ* and a thickness of 10 to 30 nm was determined by reflectometry. Such an IL film is stable under vacuum conditions. It would also only lead to additional mass of below 20 mg/m². At about 5 A/m² $\approx$ 3x10$^{19}$ e/(s·m²), a typical sample charging rate in SEM, imaging is possible with no noticeable contrast changes over many hours; this electron current density is already 6 orders of magnitudes higher than "worst case geosynchronous environments" of 3x10$^{-6}$ A/m² [1]. Measurements of the surface potential are used for further insights in the reaction of IL films to the electron beam of a SEM. Participating mechanisms such as polarization or reorientation will are discussed.

## 1. INTRODUCTION

Satellites are constantly subjected to an influx of high energetic charged particles leading to, among other effects, the build-up of differential potentials on various parts of the satellites surface, if these are not in electrical contact with each other (e.g., adjacent cover glass plates of the satellites solar cells or main body). Especially at high altitude (> 10000 km) or orbits with high inclination (> 50°) this effect is dangerous as the potential difference can exceed several thousand volts, leading to sensor malfunctions or discharges, which severely damage the satellite [2]. For example, such a discharge did severe damage to the solar array of the ESA EURECA mission [2].

Common solutions to mitigate the dangers of differential charging include coating the cover glass of the solar cells with indium tin oxide or design adaptations such as increasing the distance between individual solar cells. However, these come with their own disadvantages, mainly high cost and increased payload mass.

An ideal solution would be a conductive coating for the cover glass, that is inexpensive, transparent for visible light and parts of the infrared spectrum, easy to apply, stable under space conditions and would not lead to a significant mass increase. A class of materials that could satisfy these conditions are ionic liquids, thus in this work we explore how thin films can be applied to glass surfaces and how those films behave under ultra-high vacuum when subjected to the electron beam of a scanning electron microscope (SEM), mimicking the conditions in space.

## 2. SAMPLE PREPARATION

We used conventional glass cover slides (Menzel) as substrates. In order to have a reference structure for microscopy, gold nano-triangles were prepared on the surface of the samples using nanosphere lithography as developed by Fischer and Zingsheim [3]. In brief, 100 µl of a 50% aqueous suspension of polystyrene (PS) micro or nanospheres (microparticles GmbH) were drop-casted on the substrate, resulting in a hexagonally arranged layer of PS spheres. Subsequently, a gold film of 30 nm (as indicated by a quartz microbalance) was evaporated on the PS sphere. These spheres, including their gold coverage, were then removed using tetrahydrofuran, leaving only the gold behind, that was deposited directly onto the glass. This results in an array of approximately triangular metal islands with occasional defects. By varying the diameter of the PS spheres used in this procedure, edge lengths of the triangles can be tuned from 1.5 µm to 150 nm. The thickness of the triangles (~30 nm) has been verified by atomic force microscopy. Afterwards the samples were UV-ozone cleaned (PSD series, Novascan) for two hours to remove any residual organic substances. Furthermore, the ozone treatment leads to an increased hydrophilicity of the surface, making it easier to prepare uniform films of polar or polarizable species. After the ozone treatment, a solution of 100 µl deionized water containing 3% of the ionic liquid 1-Butyl-1-methylpyrrolidinium dicyanamide (BMP DCA, Iolitec) was pipetted onto the surface. After 1 minute the excess liquid was removed by wiping it using a lab wipe. Finally, the sample was left to dry in air.

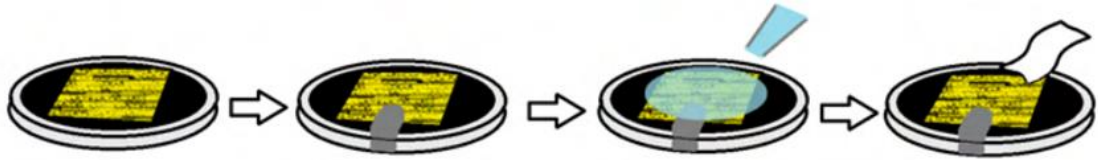

Fig. 2: Scheme of the Ionic liquid thin layer preparation

## 3. COATING THICKNESS

### 3.1 Reflectometry

The thickness of the BMP DCA thin film was determined by reflectometry using a commercial device (NanoCalc-XR, Ocean Optics). For this, a sample was placed on a silicon substrate and illuminated with a deuterium lamp. The resulting spectrum was fitted, assuming a stacked system made of BMP DCA, glass (n=1.46 [4]) and silicon (n=3.98 [5]) (from top to bottom), using the known refractive indices of glass, silicon and BMP DCA (n=1.5 [6]). This yielded a thickness for the BMP DCA layer of 12.8 ±0.8 nm. Such layers would increase the mass of solar panels by about 15 mg/m².

## 3.2 AFM measurements

To confirm that the BMP DCA forms a homogeneous layer on the substrate, additional ex situ AFM analyses were performed in which a glass sample with gold nanotriangles was compared to a glass sample with gold nanotriangles and an additional coating of BMP DCA. Measurements were carried out in a commercial instrument (Park Systems XE-100) using non-contact silicon cantilevers (SSS-NCHR-50, Nanosensors), in dynamic mode.

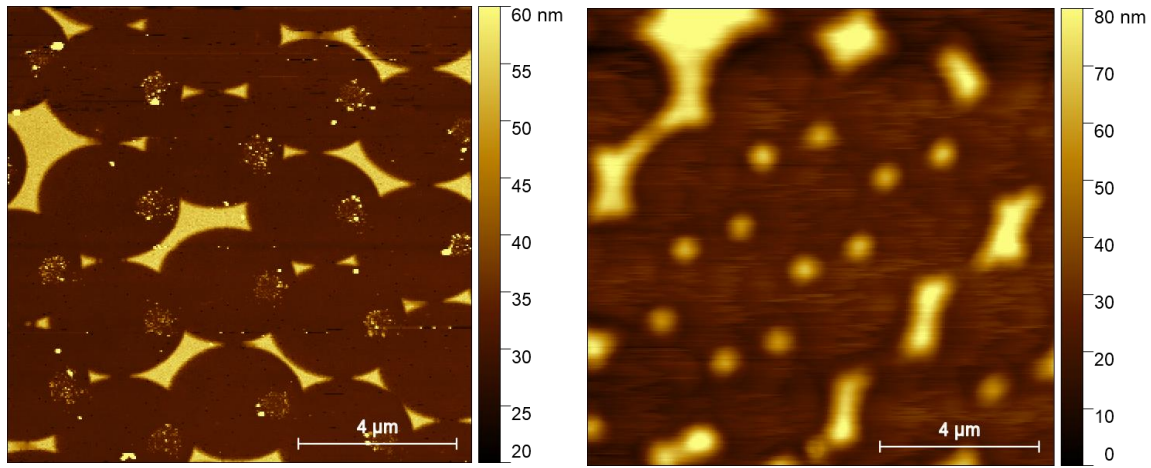

Fig. 3: AFM topography images of Au nanostructures prepared by nanosphere lithography (using 3 µm spheres as mask) without (left) and with (right) a coating of BMP DCA. In this defect-rich region the effect on various shapes and sizes of Au islands is evident.

Figure 3 shows the AFM topography images of these uncoated (left) and coated (right) samples. For the uncoated sample, the gold structures are clearly visible with sharp edges and exhibiting heights of ~25 nm. Between the gold structures residual material from the assembled polystyrene spheres is visible. As this residue is resistant to UV-ozone treatment, it probably is inorganic material. This material seems to be completely absent on the sample with BMP DCA coating, either because the application of the liquid and successive wiping of the sample removed it, or because the BMP DCA film now covers it and as the films surface is imaged it appears to be absent.

The gold structures on the coated material have far less pronounced edges compared to the uncoated sample. Because of the repulsive interaction of the BMP DCA layer with silicon and the oscillation frequency of ~350 kHz of the cantilever, the periodic formation of a meniscus of liquid between tip and samples should be repressed [7]. Thus, the surface of the liquid can be imaged at the expense of resolution. This is supported by the observation of droplet hills on the gold. The height of the gold structures with respect to the glass substrate seems to have increased to up to 60 nm. Also, it no longer forms plateaus. The formation of these droplets can be explained by the BMP DCA forming menisci at the edges of the gold triangles, effectively smoothing them out. This points to the formation of curved droplets on top of the gold triangles with an increased ionic liquid layer thickness around them. However, as the general shape of the gold structures is still recognizable, the layer thickness must be on a scale of several tens of nanometers, in good agreement with the previous reflectometry measurements. As realistic cover glass for solar cells does not require the preparation of nanostructures, this effect is expected to absent here.

## 4. SEM IMAGING

The idea to use ionic liquid as a coating for insulating materials to be able to image them in an electron microscope has been tested in literature before [8,9]. However, those samples were usually biologic samples such as cells or pollen, not flat samples such as glass, which require a thin and uniform film. Consequently, the conductive properties of such films were, to our knowledge, not yet tested in a SEM. To simulate the charging behavior in space environments, the samples were mounted on a metal carrier using conductive carbon pads and exposed to the electron beam of a scanning electron microscope (EVO MA 10, Zeiss) at high vacuum conditions ($10^{-6}$ mbar). At beam currents of ~100 pA, the current density of 5 A/m$^2$ ≈ 3×10$^{19}$ e/(s·m$^2$) is approximately six order magnitudes higher than the 6×10$^{12}$ e/(s·m$^2$) in average and 1.66×10$^{13}$ e/(s·m$^2$) under storm conditions that have been detected by ATS and SCATHA missions [1]. Hence a material that is capable of mitigating charging effects at SEM current densities should be easily capable to perform that task under real space conditions.

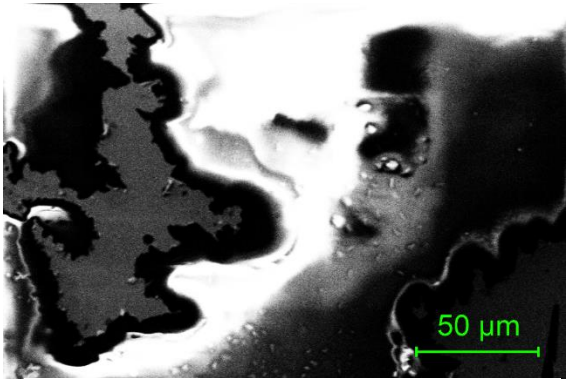
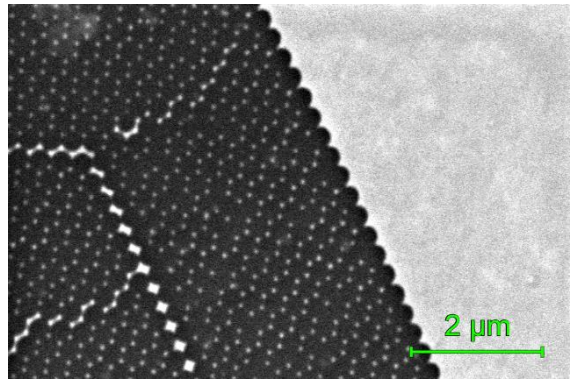

Fig. 4a: SEM image of bare glass with metal nanostructures, 5keV, 2.8x10$^{12}$ e/m$^2$

Fig. 4b: SEM image of BMP DCA treated glass with metal nanostructures, 5keV, 1.5x10$^{18}$ e/m$^2$

Figures 4a and 4b show a comparison between a glass sample with gold nanotriangles (4a) and a glass sample with gold nanotriangles and an additional coating of BMP DCA (4b), both subjected to the aforementioned conditions. When imaging the uncoated sample using a primary electron energy 5 keV, no clear image of the surface could be achieved, even at comparatively low magnification, i.e., low electron densities of 2.8×10$^{12}$ e/m$^2$. The primary feature visible is a bright region in the center, which is the result of charges accumulating on the surface and reflecting the primary electrons towards the detector [10]. Those bright areas are surrounded by dark band like features which are the result of secondary electrons emitted from the surface being guided away due lateral fields, which are a result of the charges on surface in the bright regions [10].

On the coated sample however, the surface of the sample with its gold nanostructures, primarily the triangles with edge lengths of below 100 nm, is clearly visible. The same primary electron energy as before was used, however, at a far higher magnification corresponding to almost six orders of magnitude higher electron density of 1.5×10$^{18}$ e/m$^2$.

The gold appears brighter than the glass, because of its higher electron density, leading to an increased cross section and emission of secondary electrons. In the top left figure 4b a brighter cloud like feature can be seen, which is the onset of surface charging, which may be explained by a not entirely uniform film. Upon further increasing the magnification (thus the electron current density), these regions with possibly thinner films give rise to features like the one in 4a, albeit on a much smaller scale. However, as mentioned above, these are extreme conditions, unlikely to occur in space.

## 5. SURFACE POTENTIAL MEASUREMENTS

To better understand the conduction mechanism of BMP DCA, measurements of the surface potential were performed. These mechanisms could be that the electrons are conducted by "hopping" from one molecule to the other and thus the charging described in section 4 is actually the limit of this conduction channel or perhaps the electron beam might induce some irreversible chemical reactions, to name a few. In a first step glass samples containing gold nanostructures were compared to samples containing the nanostructures and a BMP DCA film again. The measurements were performed in the UHV chamber (~$10^{-10}$ mbar) of a tuning fork based scanning probe microscope (RHK, Duoprobe), equipped with a UHV compatible SEM (Orsay Eclipse+). Such a device detects the change of the eigenfrequency $df$ of a tuning fork, when the tip attached to it interacts with a surface. This change is proportional to the force gradient with $z$ denoting the tip-sample-separation, as described in eq. (1). Assuming a model in which the tip and the sample form capacitor, the force $F$ resulting from an applied voltage $V$ can be described by eq. (2), with $A(z)$ denoting a prefactor dependent on the tip-sample geometry.

$$(1) \quad -\frac{dF}{dz} \propto df \qquad (2) \quad F = A(z)V^2$$

If the applied voltage compensates the potential on the surface, the force acting on the tip and thus the frequency shift become minimal. The observed surface potential can directly be attributed to the presence of charges on the surface.

Figure 5 shows such a surface potential measurement for glass with gold nanostructures (red) and glass with gold nanostructures and BMP DCA coating (blue) after exposure to ~$10^{17}$ e/m² each. The measured frequency shift is plotted versus applied voltage. For the measurement on uncoated glass, the maximum of the curve is clearly shifted towards a positive voltage. As a positive voltage is applied, this means the actual potential of the surface is negative, which is what one would expect for a negatively charged surface. Fitting the curve, a minimal frequency shift is reached at 33.6 V bias Voltage, which means the surface potential is -33.6 V.

For the coated sample, on the other hand, the fit parabola has its maximum at +0.22 V bias voltage and thus surface potential of -0.22 V is observed. This small surface potential is not necessarily caused by the presence of charges but rather by the contact potential difference of tip and sample.

Note that the different sign in the frequency shift also has a physical reason. As the charges on the surface induce image charges in the tip and both attract each other, for uncoated glass attractive forces are dominant (as long as the tip does not directly touch the sample in which case Pauli-repulsion becomes dominant), which result in negative frequency shifts. For the coated glass, these surface charges are absent meaning the interaction between tip and sample follows a general Lennard-Jones like behavior with attractive and repulsive regimes. Approaching in this repulsive regime corresponds to the positive frequency shift observed.

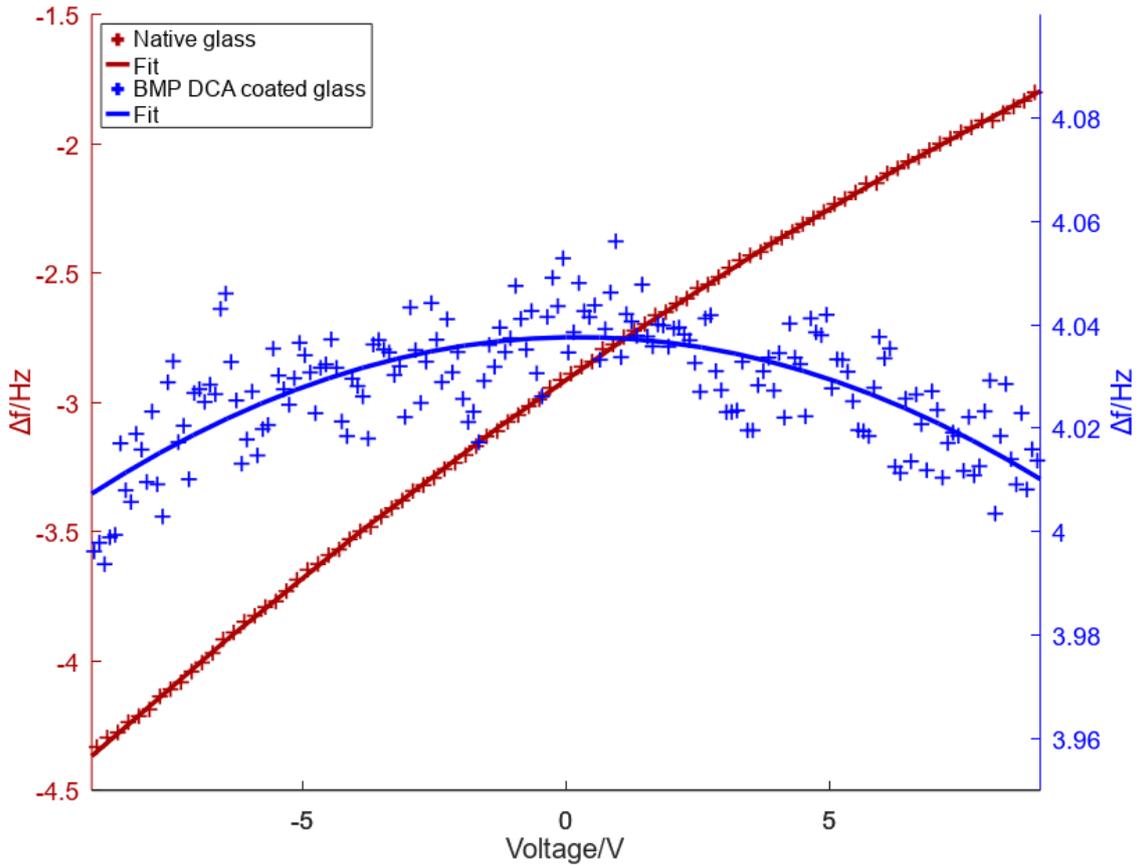

Fig. 5: df-V-spectroscopy curves for glass with gold nanostructures (red) and glass with gold nanostructures and BMP DCA coating (blue) after exposure to ~$10^{17}$ e/m² each. Tip material: Pt/Ir
Crosses indicate experimental data, continuous lines the parabolic least-square fits
Red fit: -0.00213 Hz/V² × (V-33.6 V)² - 0.51 Hz; blue fit: -0.00036 Hz/V² × (V-0.22 V)² + 4.04 Hz

## 6. CONCLUSIONS

BMP DCA can be used to generate nanoscopically thin conductive layers on glass substrates, which was confirmed by AFM and reflectometry measurements. By imaging the gold nanotriangles which were generated on top of the glass surface in a scanning electron microscope, we confirmed that these layers were indeed sufficient to compensate charge densities that are far beyond typical conditions encountered in space. Furthermore, measurements of the surface potential of BMP DCA coated surfaces indicated the absence of residual surface charges, even after exposure to comparatively high electron densities.

This suggests, that BMP DCA coatings should be capable to be used as a coating to mitigate spacecraft surface charging.

## 7. FUTURE WORK

In the future we plan to investigate, how these ionic liquid coatings behave under the influence of ionizing radiation, i.e., whether they retain their conductive capabilities, whether uniform films are damaged and whether their optical properties change. Studies indicate almost no absorbance above 250 nm [11]. Additionally, time and spatially resolved Kelvin Probe Force microscopy measurements of the surface potential seem promising for unravelling the nature of conduction mechanisms in ionic liquids. Ultimately, a performance test of these layers under real space conditions will be necessary.